\documentclass[submission,copyright,creativecommons]{eptcs}

\providecommand{\event}{Linearity 2012}                                 \usepackage[T1]{fontenc}
\usepackage{amsmath,amsfonts,amssymb,stmaryrd,amsthm}
\usepackage{listings}
\usepackage{xspace}
\usepackage{cmll}
\usepackage{hyperref}
\usepackage{eurosym}

\newtheorem{theorem}{Theorem}\newtheorem{lemma}[theorem]{Lemma}

\newtheorem{corollary}[theorem]{Corollary}

\newcommand{\grmeq}{\; ::= \;}
\newcommand{\keyword}[1]{\textsf{\upshape\small #1}\xspace}

\newcommand{\subs}[2]{[{#1}/{#2}]}
\newcommand{\rulename}[1]{\text{\sc #1}\xspace}
\newcommand{\typeconst}[1]{\keyword{#1}}

\newcommand{\tensor}{\mathbin{\otimes}}

\newcommand{\fv}{\operatorname{fv}}      
\newcommand{\one}{\textbf{1}}

\newcommand\unit{()}
\newcommand\out{!}
\newcommand\inp{?}
\newcommand{\assume}{\keyword{assume}\;}
\newcommand{\assert}{\keyword{assert}\;}
\newcommand{\new}[1]{(\keyword{new}\; #1)}
\newcommand{\PAR}{\mid}
\newcommand{\REP}{*}
\newcommand\END{\mathbf{0}}

\newcommand{\un}{\keyword{un}}
\newcommand{\lin}{\keyword{lin}}

\newcommand{\heat}{\Rrightarrow}

\newcommand{\reductionrulename}[1]{\rulename{R-{#1}}}
\newcommand{\RuleCom}{\reductionrulename{Com}}
\newcommand{\RuleAssume}{\reductionrulename{Assume}}
\newcommand{\RuleAssert}{\reductionrulename{Assert}} 
\newcommand{\RulePar}{\reductionrulename{Par}} 
\newcommand{\RuleRes}{\reductionrulename{Res}} 
\newcommand{\RuleStruct}{\reductionrulename{Heat}} 
\newcommand{\osred}{\,\rightarrow\,}

\newcommand{\REC}{\mu}
\newcommand\Tunit{\typeconst{unit}}
\newcommand{\typerulename}[1]{\rulename{T-{#1}}}
\newcommand{\nat}{\typeconst{nat}}
\newcommand{\TEND}{\typeconst{end}}
\newcommand{\TRulePar}{\typerulename{Par}}
\newcommand{\TRuleEnd}{\typerulename{End}}
\newcommand{\TRuleRes}{\typerulename{Res}}
\newcommand{\TRuleUnit}{\typerulename{Unit}}
\newcommand{\TRuleVar}{\typerulename{Var}}
\newcommand{\TRuleOut}{\typerulename{Out}}
\newcommand{\TRuleIn}{\typerulename{In}}
\newcommand{\TRuleAssume}{\typerulename{Assume}}
\newcommand{\TRuleAssert}{\typerulename{Assert}}
\newcommand{\TRuleRep}{\typerulename{Rep}}

\newcommand{\TRuleConv}{\typerulename{Conv}}
\newcommand{\TRuleTensorE}{\typerulename{$\otimes$E}}
\newcommand{\TRuleOneE}{\typerulename{$\one$E}}
\newcommand{\TRuleRefE}{\typerulename{RefE}}
\newcommand{\TRuleForm}{\typerulename{Form}}
\newcommand{\TRuleOneI }{\typerulename{$\one$I}}
\newcommand{\TRuleTensorI}{\typerulename{$\otimes$I}}
\newcommand{\TRuleRefn}{\typerulename{RefI}}

\newcommand{\dom}{\operatorname{dom}}
\newcommand{\csplit}{\circ}
\newcommand{\tp}{\colon}
\newcommand{\cupdate}{+}

\newcommand{\der}{\vdash}
\newcommand{\derwf}{\vdash_{\textsf{\upshape{\tiny\!wf}}}}

\newcommand{\dual}[1]{\overline{#1}}

\newcommand{\cf}{\operatorname{cf}}      

\newcommand{\Client}{\textbf{Client}}
\newcommand{\Store}{\textbf{Store}}
\newcommand{\Bank}{\textbf{Bank}}
\newcommand{\Charge}{\textbf{Charge}}
\newcommand{\product}{\typeconst{product}}
\newcommand{\ccard}{\typeconst{ccard}}

\title{Linearly Refined Session Types}
\author{Pedro Baltazar \qquad\qquad Dimitris Mostrous \qquad\qquad Vasco T. Vasconcelos
  \institute{University of Lisbon, Faculty of Sciences and LaSIGE\\ Lisbon, Portugal}
  \email{\{pbtz,dimitris,vv\}@di.fc.ul.pt}
}
\def\titlerunning{Session Types With Linear Refinements}
\def\authorrunning{P. Baltazar, D. Mostrous \& V. T. Vasconcelos}

\begin{document}
\maketitle

\begin{abstract}
  Session types capture precise protocol structure in concurrent
  programming, but do not specify properties of the exchanged values
  beyond their basic type.  Refinement types are a form of dependent
  types that can address this limitation, combining types with logical
  formulae that may refer to program values and can constrain types
  using arbitrary predicates.  We present a pi calculus with assume
  and assert operations, typed using a session discipline that
  incorporates refinement formulae written in a fragment of
  Multiplicative Linear Logic.  Our original combination of session
  and refinement types, together with the well established benefits of
  linearity, allows very fine-grained specifications of communication
  protocols in which refinement formulae are treated as logical
  resources rather than persistent truths.
\end{abstract}

\section{Introduction}
\label{sec:introduction}

\emph{Session types}~\cite{honda.vasconcelos.ea:language-primitives}
are a practical and expressive type-based verification methodology for
concurrent programming, and have proved excellent in modeling typed
computations predominantly consisting of client-server message
passing.  As a simple example, assigning the type $\oc
\mathsf{int}.\wn \mathsf{bool}.\mathsf{end}$ to a communication
channel means that a value of type $\mathsf{int}$ will be sent ($\oc
\mathsf{int}$), then a $\mathsf{bool}$ will be received ($\wn
\mathsf{bool}$), and the channel cannot be used any further.
Communication soundness is ensured when the ``other end'' of the
communication channel is used in a complementary (or dual) way:
$\overline{\oc \mathsf{int}.\wn \mathsf{bool}.\mathsf{end}} = \wn
\mathsf{int}.\oc \mathsf{bool}.\mathsf{end}$.

\emph{Refinement types}, as defined for ML~\cite{Freeman:1991:ML}, are
a form of dependent types that allow the programmer to attach formulae
to types, thus narrowing down the set of values inhabiting a given
type.  For instance, the type $\{ x : \mathsf{int} \: | \: 0 \le x
\wedge x \leq 10 \}$ describes integer values in the range $0..10$.
Such fine-grained types have met increasing attention, with several
notable works on type checking for functional programming, such as
hybrid type checking~\cite{Flanagan:2006:HTC}, liquid
types~\cite{Rondon:2008:LT}, or the blame
calculus~\cite{wadler.findler:well-typed-programs-not-blamed}.  In the
context of this work, let us note that refinements for ML written in
Intuitionistic Linear Logic have been introduced
in~\cite{mandelbaum:effref}.
A detailed overview is
in~\cite{gordon.fournet:principles-refinement-types}.

With regard to refinement formulae, the most common approach is to use
classical first-order logic, which is certainly enough for many
examples, but cannot provide a satisfactory treatment of refinements
on resources. In particular, it does not allow one to control finer
computational properties: a type $\{ x\colon \mathsf{ccard} \: | \:
\mathsf{use}(x) \}$ may mean that we can use a credit card, but it
does not mandate that we can do so \emph{just once}.  To achieve such
finer distinctions between types, we specify refinements in a fragment
of {\em multiplicative linear logic} (MLL)~\cite{Girard87}, most
notably without exponentials or additives.

Building on previous work on session
types~\cite{vasconcelos:fundamental-sessions}, we combine sessions and
linear refinements to obtain an original system of \emph{linearly
  refined session types}, noting that until now neither linear nor
classical refinements have been studied in the context of session
types, according to our knowledge.  The result is a system in which
typed message exchange, refinement, and resources are combined,
providing for a very fine control of \emph{process behaviour}.  We
show that well-typed programs do not get stuck when trying to verify
logical properties.

The rest of the paper is structured as follows.
Section~\ref{sec:language} introduces the language and its operational
semantics, as well as the running example. Section~\ref{sec:typing}
describes the typing system and Section~\ref{sec:results} the main
results.  We conclude the paper with related work and future
directions.

\section{The pi calculus with assume and assert}
\label{sec:language}

Consider a simple online Store that accepts a product request from a
Client, and interacts with a Bank to perform the payment. The Store
and the Client share a private channel in which the Client sends the
product $\text p$, the credit card number $\text c$ and the price it
is willing to pay, \euro100.  The Store acts dually by accepting the
product, the credit card, and the amount to be charged, and by
immediately charging, using $\Charge(c,a)$, the amount $a$ to the
credit card $c$.
\begin{equation*}
  \Client =  s_1 \out \text{p} . s_1 \out \text{c} . s_1 \out 100 . \END 
  \qquad
  \Store = s_2 \inp p . s_2 \inp c . s_2 \inp a . \Charge(c,a)
\end{equation*}
In the code above, $s_1 \out \text{p}$ means to send the value $\text
p$ on channel endpoint $s_1$, dually $s_2 \inp p$ means to read a
value from $s_2$ binding it to variable $p$, and $\END$ denotes the
terminated process. Value $\text p$ should not be confused with
variable $p$. In our language processes read and write within sessions
by using distinct variables to identify the two ends of the channel,
$s_1$ and $s_2$ in this case.

In order to charge the Client, the Store calls the Bank service, and sends the credit card number and the amount to be
charged.
\begin{equation*}
  \Bank = \REP r_1 \inp y . y \inp c . y \inp a . \END
  \qquad
  \Charge(c,a) = \new {b_1 b_2} (r_2 \out b_2 . b_1 \out c . b_1 \out a.\END) 
\end{equation*}
In the Bank code, $\REP$ prefixes a replicated process that can be
used an unbounded number of times, as one would expect in this
example.  The \Charge{} process creates a new channel with the $\new
{b_1 b_2}$ constructor, whose purpose is to establish a private,
bidirectional channel with the bank.  To set up the session, the
channel endpoint $b_2$ is passed to the bank and the other, $b_1$, is
retained locally for interaction with the bank.  Note that the
language is explicitly typed, but for brevity we ignore the type
annotations in our examples.

The overall system is the parallel composition of the three processes
connected by two channels: $r_1 r_2$, the public Bank-Store channel,
and $s_1 s_2$, the private Client-Store channel.
\begin{equation*}
  \new{r_1 r_2}\new {s_1s_2} (\Client \PAR \Store \PAR \Bank )
\end{equation*}

\begin{figure}[t]
  \begin{align*}
    \varphi  \grmeq     && \textit{Formulae:} &\qquad\qquad   &  P  \grmeq  && \textit{Processes:}\\
    & A(v_1,\ldots,v_n)  & \text{predicate on $v_1,\ldots,v_n$}  &&   & x\out v.P & \text{output}\\
    & \varphi \tensor \varphi & \text{joining}       &&                            & x\inp x.P & \text{input}\\
    & \one  &  \text{identity}                   &&                            & P\PAR P & \text{parallel composition}\\
    & &								&&							& \REP P & \text{replication}\\	
    v \grmeq & & \textit{Values:}         &&                            & \END & \text{inaction}\\
    &  x  & \text{variable}               &&                            & \new{xx \tp T}P & \text{scope restriction}\\
    &  \unit & \text{unit}                &&	                        &  (\assume \varphi) P & \text{assume} \\
    &&                                    &&                            & \assert \varphi.P & \text{assert} 
  \end{align*}
  \caption{The syntax of processes}
  \label{fig:syntax-processes}
\end{figure}

The syntax of processes is presented in
Figure~\ref{fig:syntax-processes}. The linear nature of sessions, for example in the session $s_1 s_2$ between Client and Store, can
ensure some security properties. By enriching such a calculus with cryptography primitives, more
properties can be captured, such as authentication requirements and
privacy of communication (e.g.~\cite{Bhargavan:2010:MVS}).  However,
even if such properties are satisfied the system can contain
unintended uses of given permissions by authorized processes.
In the above example, the Store can wrongly compute the amount to be
charged, which will be detected only later by the Client.
\begin{equation*}
  \Store_{\textbf 1} = s_2 \inp p . s_2 \inp c . s_2 \inp a  . \Charge(c,a+10)
\end{equation*}
A more subtle situation is when two threads try to charge the Client for the same
purchase.
\begin{equation*}
  \Store_{\textbf 2} = s_2 \inp p . s_2 \inp c . s_2 \inp a . (\Charge(c,a) \PAR \Charge(c,a))
\end{equation*}

Our language enriches pi calculus with $\assume$ and $\assert$
commands, using formulae $\varphi$ built over a set of uninterpreted
predicates $A,A_1,A_2,\ldots$, the linear logic connective of
tensor, $\tensor$, and its identity,~$\one$. The predicates may refer
to channel names or base-values such as integers and strings, which
are represented here by the unit value, $\unit$; therefore,
refinements form dependent types.
Enhanced with these commands, the Client may assume a $charge(\text
c,100)$ capability on the values sent to the Store.  And the Bank, in
turn, will assert that exact capability.
\begin{gather*}
  \Client_{\textbf 1} = (\assume charge(\text c,100)) \textbf{Client}
  \qquad
  \textbf{Bank}_{\textbf 1} = \REP r_1 \inp y . y
  \inp c . y \inp a . \assert charge(c,a).\END
\end{gather*}

\begin{figure}[t]
  \emph{Heating relation}, $P\heat Q$ \qquad($P\equiv Q$ means $P\heat
  Q$ and $Q\heat P$)
    \begin{gather*}
    P\PAR Q \equiv Q\PAR P
    \qquad
    (P\PAR Q)\PAR R \equiv P\PAR (Q\PAR R)
    \qquad
    P\PAR \END\equiv P
    \qquad
    \REP P \equiv P \PAR \REP P
    \qquad
    \new{xy \tp T}\END \equiv \END
    \qquad
    \\
    \new{xy \tp T} (P \PAR Q)  \equiv  \new{xy \tp T} P \PAR Q       \qquad
    \new{wz \tp T}\new{xy \tp U}P\equiv\new{{xy \tp U}}\new{wz \tp T}P
    \\
    \new{xy \tp T}(\assume \varphi) P \equiv (\assume \varphi) \new{xy \tp T} P 
            \qquad
    (\assume\one) P \equiv P
    \qquad
    \assert\one. P \equiv P
    \\
    (\assume{\varphi_1})(\assume{\varphi_2})P \equiv (\assume{\varphi_2})(\assume{\varphi_1})P 
    \qquad
    \assert \varphi_1. \assert \varphi_2. P \equiv \assert \varphi_2. \assert \varphi_1. P
    \\
    \assert \varphi_1 \tensor \varphi_2.P \equiv \assert \varphi_1. \assert \varphi_2. P
    \qquad
    (\assume \varphi_1 \tensor \varphi_2 )P \equiv (\assume \varphi_1)
    (\assume \varphi_2) P
    \\
    (\assume\varphi)P\PAR Q \heat (\assume\varphi)(P\PAR Q) 
    \qquad
    \new{xy \tp T} P   \equiv  \new{xy \tp U} P   \quad \text{ if } T \equiv U
  \end{gather*}
    \emph{Reduction relation}, $P\to Q$
    \begin{gather*}
    \tag\RuleCom
    \new{xy \tp (q ! w \tp T . U) }(x \out v . P \PAR y \inp z. Q \PAR R)
    \osred
    \new{xy \tp U\subs v w}(P \PAR Q \subs v z \PAR R)
    \\
    \tag{\RuleAssert} 
    (\assume \varphi)(\assert {\varphi} . P \PAR Q) \osred P \PAR Q
    \\
    \tag{\RuleAssume, \RuleRes} 
    \frac
    {P \osred Q}
    {(\assume \varphi)P \osred  (\assume \varphi) Q}
    \qquad \qquad 
    \frac
    {P \osred  Q}
    {\new{xy \tp T}P \osred \new{xy \tp T}Q}
    \\
    \tag{\RulePar, \RuleStruct}
    \frac
    {P \osred  Q}
    {P \PAR R \osred Q \PAR R}
    \qquad \qquad 
    \frac
    {P \heat P' \qquad P' \osred Q' \qquad Q' \heat Q}
    {P \osred Q}
  \end{gather*}
  \caption{Operational semantics}
  \label{fig:HeatP}
\end{figure}

In order to explain the interplay between \keyword{assume} and
\keyword{assert}, we turn our attention to the operational semantics
of the language.
We say that variable $y$ occurs \emph{bound} in process $P$ within
$x\inp y.P$ and $\new{xy \tp T}P$, in type $U$ within $q?y\colon T.U$
and $q!y\colon T.U$, and in formula $\varphi$ within $\{y\colon T |
\varphi \}$. Also, variable $x$ occurs bound in $\new{xy \tp T}P$. A
variable that occurs in a non-bound position within a process, type, or formula is said to be \emph{free}. The sets of free variables in a
process $P$, a type $T$ or a formula~$\varphi$, denoted by $\fv(P)$,
$\fv(T)$ and $\fv(\varphi)$, are defined accordingly and so is
alpha-conversion.
We work up to alpha-conversion and follow Barendregt's variable
convention, whereby all variables in binding occurrences in any
mathematical context are pairwise distinct and distinct from the free
variables.

The standard capture-free \emph{substitution} of variable~$x$ by
value~$v$ in process~$P$, a type $T$ or a formula $\varphi$, is
denoted by $P\subs vx$, $\varphi\subs vx$ and $T\subs vx$.
This follows the standard treatment for dependent session/channel
types~\cite{mostrous.yoshida:two-session-systems-higher-order,
  Yoshida:2004:CDT}.
For example, the substitution $(\new{xy\tp T} P) \subs v z$ is defined
as $\new{xy\tp T\subs v z} P\subs v z$.

From the operational semantics we factor out a \emph{heating relation}
meant to simplify the statement of the reduction relation, by structurally
adjusting processes. Both relations, heating and reduction, are
defined in Figure~\ref{fig:HeatP}. We start with reduction.
The relation includes the rule for communication, \RuleCom, adapted 
from~\cite{vasconcelos:fundamental-sessions} to handle dependent refinements, 
and the usual rules for
reduction underneath parallel composition and restriction, \RulePar,
\RuleRes, and under 
heating with \RuleStruct. It also includes two
novelties: an axiom \RuleAssert for cutting assertions, and a rule
that allows reduction under assumptions, \RuleAssume.
The correspondence of \RuleAssert with the logical cut is evident,
noting that a choice has been made for assumptions to define a scope
and for the cut to take place against enclosed assertions. The
alternative would be for the cut to take place between an
\keyword{assume} and an \keyword{assert} in parallel, but at the
typing level this would require a form of negation which would
effectively identify assumptions and assertions; instead of asserting
$\varphi$ one could assume $\varphi^\bot$ and the two possibilities
would be indistinguishable at the typing environment level.  As a
result, two assumes could cancel out each other, and similarly for two
asserts, thus compromising the intended usage of assertions.

On what concerns heating, the rules in the first two lines are
standard in the pi calculus, those in the following three lines
manipulate \keyword{assume} and \keyword{assert} processes, as well as
linear logic formulae, in the expected way. The last line introduces
the only truly directional rule, allowing the scope of an assumption
to encompass another process.
The reason why the rule is not bidirectional is because we want to
keep assertions in the scope of assumptions; take for example a
process $P$ of the form $(\assume A)(\END \PAR
\assert A.\END)$.  We have that $P$ reduces in one step to $\END$, but
$(\assume A)\END \PAR \assert A.\END$ is stuck.
With assume, and unlike scope extrusion, i.e., $\new{xy} P\PAR Q
\equiv \new{xy}(P\PAR Q)$, we do not have bound variables to control
the application of the rule.
The last rule in the figure allows to expand a recursive type, paving
the way applications of rule \RuleCom.
Notice that we do not mention the usual sideconditions, e.g., that
$x,y~\notin~\fv~(Q)$ in the scope extrusion rule, since the variable
convention can be assumed to provide this guarantee.

In the example, by heating, the $(\assume charge(\text c,100))$ can be
extended to encompass the \Store{} process, and then moved to a
position before session creation $\new{s_1s_2}$ to allow the
interaction between the Client and the Store on channel $s_1s_2$, via
the \RuleCom rule.
\begin{multline*}
  \new{s_1s_2}(\Client_{\textbf 1} \PAR \Store) \heat
  \\
  (\assume charge(\text c,100))\new{s_1s_2}( s_1 \out \text p . s_1 \out
  \text c . s_1
  \out 100 . \END \PAR s_2 \inp p . s_2 \inp c . s_2 \inp a
  . \Charge(c,a)) \to \to \to
  \\
  (\assume charge(\text c,100)) \Charge(\text c,100)
\end{multline*}
Next, the process is ready to perform the communication between
$\Bank_{\textbf 1}$ and \Store. Rule {\RuleAssert} matches the assume
with the assert, and the process is concluded.
\begin{multline*}
  \new{r_1r_2}(\assume charge(\text c,100))(
  \Charge(\text c,100) \PAR \Bank_{\textbf 1}) \equiv
  \\
  (\assume charge(\text c,100)) \new{r_1r_2}  (
  \Charge(\text c,100)
  \PAR r_1 \inp y . y \inp c . y \inp a . \assert charge(c,a) 
  \PAR\Bank_{\textbf 1} )
  \to \to \to
  \\
  \assume charge(\text c,100)\, \assert charge(\text
  c,100).\END
  \PAR
  \new{r_1r_2}\Bank_{\textsf 1}
  \to
    \new{r_1r_2} \Bank_{\textbf 1}
\end{multline*}

Clearly, if $\Store_{\textbf 1}$ is used, the reduction will yield a
process where the assumption and the assertion do not match.
\begin{multline*}
  \new{r_1r_2}\new{s_1s_2} (\Client \PAR \Store_{\textbf 1} \PAR \Bank_{\textbf 1})
  \to \cdots \to
  \\
  \assume charge(c,100)\,\assert charge(c,110).\END
  \PAR
  \new{r_1r_2} \Bank_{\textbf 1}
  \not\to
\end{multline*}

In turn, if $\Store_{\textbf 2}$ replaces $\Store_{\textbf 1}$ in the
above process, then we reach a situation where one assertion is left
unmatched.
\begin{multline*}
  \new{r_1r_2}\new{s_1s_2} (\Client \PAR \Store_{\textbf 2} \PAR \Bank_{\textbf 1})
  \to\cdots\to
  \assert charge(c,100).\END
  \PAR
  \new{r_1r_2}\Bank_{\textbf 1}
  \not\to
\end{multline*}

These two processes are stuck due to assume/assert problems --- in both
cases we find an assert for which no corresponding assume exists in the enclosing scope --- and will be identified as unsafe by the typing
system.

If somehow the client wants to be charged twice, then it 
can $\assume
\textit{charge}(\text c,100)$, twice in a row. Alternatively it may
utilise a more compact variant by using joining (tensor).
\begin{equation*}
  \Client_{\textbf 2} =  (\assume \textit{charge}(\text c,100)\tensor
  \textit{charge}(\text c,100)) s_1
  \out \text p . s_1 \out \text c . s_1 \out 100 . \END
\end{equation*}
Then, by taking advantage of the heating rule that allows breaking the
$(\otimes)$, as well as reduction underneath assumptions, we can easily
see that:
\begin{equation*}
  \new{r_1r_2}\new{s_1s_2} (\Client_{\textbf 2} \PAR \Store_{\textbf 2} \PAR \Bank_{\textbf 1})
  \to\cdots\to 
  \new{r_1r_2} \Bank_{\textbf 1}
\end{equation*}

We conclude this section by defining what we mean by a safe
process.  
First we introduce the notion of \emph{canonical processes}. 
A process
is in canonical form if it is of the form:
\begin{equation*}
  \new{x_1y_1:T_1} \cdots \new{x_ky_k:T_k}(\assume A_1) \cdots (\assume
  A_m)(P_1 \PAR \cdots \PAR P_n) \qquad  \text{with } k,m\geq0,  n>0
\end{equation*}
and every $P_i$ is \emph{neither} a \keyword{new}, nor an
\keyword{assume} nor a parallel composition.
A simple induction on the structure of processes easily allows us to
conclude that all processes can be heated to a process in canonical
form.

Then, we say that a process $Q$ is \emph{safe} if, for all processes
$P$ in the canonical form above such that $Q \heat P$ and every $P_i$
of the form $\assert B_i.R_i$, there is a $1 \le j \le m$ such that
$B_i = A_j$.
In other words, safe processes do not get stuck at assertion points.
The next section introduces a type assignment system that guarantees
that processes typable under unrestricted contexts are safe.

Notice that each $A_i$ and $B_i$ are atomic formulae; if not, then the
heating relation may ``break'' the tensors ($\otimes$) and eliminate
the identities ($\one$), so that in the end we may match assumptions
on atomic formulae against assertions on atomic formulae.

\section{Typing system}
\label{sec:typing}

\begin{figure}[t]
  \begin{align*}
    q  \grmeq  && \textit{Qualifiers:} &&& q\,p & \text{qualified
      session}
    \\
    & \lin & \text{linear}  &&& \{x\colon T | \varphi \} &
    \text{refinement}
    \\
    & \un & \text{unrestricted} &&&  \alpha & \text{type variable}
    \\
    p \grmeq && \textit{Session types:}  &&& \REC \alpha. T & \text{recursive type}
    \\
    & ?x\colon T.T  & \text{receive}  && \Gamma \grmeq && \textit{Contexts:}
    \\
    & !x\colon T.T  & \text{send} &&& \cdot & \text{empty}
    \\
    T \grmeq && \textit{Types:} &&& \Gamma, x\colon T & \text{type assumption}
    \\
    & \Tunit & \text{unit} &&& \Gamma,\varphi & \text{formula}
    \\
    & \TEND & \text{termination}
  \end{align*}
  \caption{The syntax of types and typing contexts}
  \label{fig:types}
\end{figure}

The syntax of types is presented in Figure~\ref{fig:types}.  Let
$\product$, $\ccard$ and $\nat$ be the types of the products sold by
the store, credit cards, and natural numbers respectively (all denoted
by $\Tunit$ in the figure).  The types of the two ends $r_1r_2$ of the
Client-Store channel, and also of the Bank-Store channel $s_1s_2$, are
as follows.
\begin{align*}
  s_1 \colon&\lin ! \product. \lin ! \ccard . \lin ! \nat . \TEND
  &
  s_2 \colon& \lin ? \product. \lin ? \ccard . \lin ? \nat . \TEND
  \\
  r_1 \colon& \mu \alpha. \un ? ( \lin ? \ccard . \lin ?
  \nat.\TEND).\alpha 
  &
  r_2 \colon& \mu
  \alpha. \un !  ( \lin ?  \ccard . \lin ?  \nat.\TEND).\alpha
\end{align*}

In types, as in processes, $!$ means output and $?$ means input,
$\TEND$ denotes a channel on which no further interaction is possible,
and the $\mu$ construct is used to write recursive types.
Qualifiers $\lin$ and $\un$ are used to control the number of threads
holding references to the channel end: exactly one in the \lin case,
zero or more for the \un case. The Client-Store channel is \lin at all
times, so that a third process cannot interfere in the interaction.
The Bank-Store channel is \un at all times, so that multiple stores may connect to the bank. Such a \un channel is used to
pass a \lin channel (of type $\lin ? \ccard . \lin ?  \nat.\TEND$),
thus establishing a private channel between the Bank and the Store. In
our example, we assume that the private Client-Store channel was
created via a similar mechanism, based on some shared channel provided
by the store.
It should be easy to see that the type $\lin ! \product. \lin ! \ccard
. \lin ! \nat . \TEND$ of the $s_1$ end of the channel naturally
describes the \Client's interaction $s_1 \out \text{p} . s_1 \out
\text{c} . s_1 \out 100 . \END $, and that the type $\un !  ( \lin ?
\ccard . \lin ?  \nat.\TEND)\dots$ closely explains the \Store's
interaction $r_1 \inp y . y \inp c . y \inp a \dots$

The above typing context is correct for process $\Client \PAR
\Store \PAR \Bank$, but it remains so even if one replaces $\Store$ by
$\Store_{\textbf 1}$ or by $\Store_{\textbf 2}$, since in both 
of these
cases the usage of the channels match the prescribed behavior. 
Thus, 
traditional session types are not enough to control
and discipline the use of resources.

In order to incorporate logical information into session types, the
syntax is augmented with logical refinements, $\{ x\colon T | \varphi
\}$. Further, and in order for formulae $\varphi$ to be able to refer
to data appearing ``previously'' in types, we name the object of
communication: in type $q ?x\colon T.U$ we allow type $U$ to refer to
the value received before via variable $x$.  Types can be refined with
the exact same formulae used for asserting and assuming in
processes. For example, the types for channels $s_1$ and $r_1$ can be
logically refined in such a a way that the amount $x$ to be charged is
subject to ``permission'' $\textit{charge}(c,x)$, where $c$ denotes
the credit card number received in a previous communication.
\begin{align*}
  s_1 \colon& \lin ! p\colon \product. \lin ! c\colon \ccard . \lin !
  a\colon \{ x \colon \nat | \textit{charge}(c,x) \} . \TEND
  \\
  r_1 \colon& \mu \alpha. \un ? y \colon ( \lin ? c \colon \ccard
  . \lin ? a \colon \{x \colon \nat | \textit{charge}(c,x)
  \}.\TEND).\alpha
\end{align*}
We will get back to our running example after introducing the type
system.

For \emph{recursive types}, type variable $\alpha$ occurs bound in
type $\mu\alpha.T$. Such types are required to be \emph{contractive},
i.e., containing no subexpression of the form $\mu \alpha_1\dots\mu
\alpha_n.\alpha_1$.
We further require types not to contain subexpressions of the form
$\mu \alpha_1\dots\mu \alpha_n.\{x\colon T\mid \varphi\}$, so that the
only interesting recursive types are session types.
We leave the treatment of recursive refinement types for future work,
where they may represent the introduction of persistent formulae,
i.e., the exponentials of linear logic.
We again follow Barendregt's variable convention, this time on type
variables $\alpha$.

\emph{Type equivalence} is a central ingredient in dependent type
systems. Here we stick to a rather simple notion. The equivalence
relation of formulae is the smallest equivalence relation, denoted by
$\equiv$, containing the axioms $\varphi_1 \otimes \varphi_2 \equiv
\varphi_2 \otimes \varphi_1$ and $\varphi \otimes \one \equiv
\varphi$.
For types, we include in the equivalence relation a recursive type
$\mu\alpha.T$ and its unfolding $T\subs{\mu\alpha.T}{\alpha}$, as well
as refinement types that differ on equivalent formulae only. The
definition, omitted, is co-inductive.

\begin{figure}[t]
  \emph{The dual of a type}, $\dual T = T$
  \begin{gather*}
    \dual{q ?x\colon T.U} = q !x\colon T.{\dual{U}}
    \qquad
    \dual{q !x\colon T.U} = q ?x\colon T.{\dual{U}}
    \qquad
    \dual{\TEND} =\TEND
    \qquad
    \dual{\REC a.T} = \mu a.\dual T
    \qquad 
    \dual a = a
  \end{gather*}
     \emph{Unrestricted types and contexts}, $\un(T)$ \emph{and} $\un(\Gamma)$ 
  \begin{gather*}
    \un(\Tunit) \qquad \un(\TEND) \qquad \un(\un\,p)
    \\
        \un(\cdot) \qquad \un(\Gamma,x\colon T) \text{ if } \un(\Gamma)
    \text{ and } \un(T)
          \end{gather*}
    \emph{Well-formed formulae, $\Gamma\derwf \varphi$, well-formed types,
    $\Gamma\derwf T$, and well-formed contexts, $\derwf \Gamma$}
  \begin{gather*}
    \frac{
      \fv(\varphi) \subseteq \dom(\Gamma)
    }{
      \Gamma \derwf \varphi
    }
    \qquad\;\;
    \frac{
      \fv(T) \subseteq \dom(\Gamma)
    }{
      \Gamma \derwf T
    }
    \qquad\;\;\qquad
    \derwf \cdot
    \qquad 
    \frac{
      \derwf \Gamma \qquad \Gamma \derwf T
    }{
      \derwf \Gamma, x: T
    }
    \qquad
    \frac{
      \derwf \Gamma \qquad \Gamma \derwf \varphi
    }{
      \derwf \Gamma, \varphi
    }
  \end{gather*}
  \caption{Type duality, unrestricted predicates, and well formed predicates}
  \label{fig:well-formed}
\end{figure}

\emph{Duality} plays a central role in the theory of session
types. The two ends of a channel are supposed to be of a dual nature
at certain points in typing derivations, namely at scope restriction
$\new{xy\colon T}P$. Examples include the types for variables $s_1$ and
$s_2$, as well as those for variables $r_1$ and $r_2$ above. The
definition is in Figure~\ref{fig:well-formed}. Duality is defined only
for session types (input, output, $\TEND$, and recursion); in
particular it is undefined for refinement types in very much the way
as it is undefined for \Tunit~\cite{vasconcelos:fundamental-sessions}.

\emph{Typing contexts} are defined in Figure~\ref{fig:types} and
include type assumptions for variables, $x\colon T$, as well as
formulae $\varphi$ known to hold. The domain of a context $\Gamma$,
denoted $\dom(\Gamma)$, is defined as $\{x \mid x\colon T \in
\Gamma\}$.

Types (and contexts) can be classified as \emph{unrestricted} or
\emph{linear}; we only need the first notion; the definition
is in Figure~\ref{fig:well-formed}. Unrestricted types, denoted
$\un(T)$, are $\Tunit$, $\TEND$ and $\un\,p$ for all $p$. Unrestricted
contexts may contain unrestricted types only, in particular they
cannot contain formulae (for these are linear).

Formulae may contain program variables. Because types may include
formulae, types may contain free program variables. Formulae and types
are well formed with respect to a context if their free variables are
in the domain of the context. Contexts contain formulae and
types. Formulae and types appearing in a context must be well formed
with respect to the ``initial'' part of the context. The definitions
of well formed contexts is in Figure~\ref{fig:well-formed}. In
particular, our system does not include (implicitly or explicitly) the
exchange rule; context $x\colon \Tunit, A(x)$ is well formed but
$A(x), x\colon \Tunit$ not.

\begin{figure}[t]
  \emph{Context split}, $\Gamma = \Gamma \csplit \Gamma$
    \begin{gather*}
    \emptyset = \emptyset \csplit \emptyset
    \qquad
    \frac{
      \Gamma = \Gamma_1 \csplit \Gamma_2 \qquad \Gamma_1 \derwf \lin\, p
    }{
      \Gamma, x\colon \lin \,p = (\Gamma_1,x\colon \lin \,p) \csplit \Gamma_2      
    }
    \qquad
    \frac{
      \Gamma = \Gamma_1 \csplit \Gamma_2 \qquad \Gamma_2 \derwf \lin\, p
    }{
      \Gamma, x\colon \lin \,p = \Gamma_1 \csplit (\Gamma_2,x\colon \lin \,p)
    }
    \\
    \frac{
      \Gamma =  \Gamma_1 \csplit \Gamma_2
      \qquad
      \un(T)
    }{
      \Gamma, x\colon T = (\Gamma_1,x\colon T) \csplit (\Gamma_2,x\colon T)
    }
    \qquad
    \frac{  \Gamma = \Gamma_1 \csplit \Gamma_2 
      \qquad \Gamma_1 \derwf \varphi }
    { \Gamma, \varphi = (\Gamma_1, \varphi) \csplit \Gamma_2  }
    \qquad
    \frac{
      \Gamma = \Gamma_1 \csplit \Gamma_2
      \qquad \Gamma_2 \derwf \varphi
    }{
      \Gamma, \varphi = \Gamma_1 \csplit (\Gamma_2, \varphi)
    }
  \end{gather*}
    \emph{Context update}, $\Gamma + x\colon T = \Gamma$
  \begin{equation*}
    \frac{
      x \notin \dom(\Gamma)
      \qquad
      \Gamma \derwf T
    }{
      \Gamma + x \colon T = \Gamma, x \colon T
    }
    \qquad
    \frac{
      \un(T)
    }{
      (\Gamma, x\colon T) + x \colon T = (\Gamma, x\colon T)
    }
  \end{equation*}
                                                                                  \caption{Context split and context update}
  \label{fig:context-splitting}
\end{figure}

Central to our type system is the \emph{context split} operator that
distributes incoming formulae and linear types to one of the output
contexts while duplicating incoming unrestricted types to both the
output contexts. The definition, a straightforward extension of the
one in~\cite{vasconcelos:fundamental-sessions} that can now handle
formulae, is in Figure~\ref{fig:context-splitting}.
Formulae in contexts are handled very much like linear type
assumptions: there is one rule to ``send'' the formula (or type
assumption) to the left context and one rule to send it to the
right. There are however new assumptions, $\Gamma \derwf \varphi$ and
$\Gamma \derwf \lin p$, meant to guarantee that the output of context
splitting are well-formed contexts.
The context update operator is used to update the type of a
channel, after its prefix has been used. It is used in the typing rules
for input and output processes.

\begin{figure}[t]
  \emph{Typing rules for formulae}, $\Gamma \der \varphi$
    \begin{gather*}
    \tag{\TRuleOneI,\TRuleForm,\TRuleTensorI}
    \frac{
      \derwf \Gamma
      \quad
      \un(\Gamma)
    }{
      \Gamma\der \one
    }
    \qquad 
    \frac{
      \derwf \Gamma_1,\varphi,\Gamma_2
      \quad
      \un(\Gamma_1,\Gamma_2) 
    }{
      \Gamma_1,\varphi,\Gamma_2 \der \varphi
    }
    \qquad
    \frac{
      \Gamma_1 \der \varphi_1
      \quad
      \Gamma_2 \der \varphi_2
    }{
      \Gamma_1 \csplit \Gamma_2 \der \varphi_1 \tensor \varphi_2
    }
   \end{gather*}
  \emph{Typing rules for values}, $\Gamma \der v:T$
    \begin{gather*}
    \tag{\TRuleUnit,\TRuleVar,\TRuleRefn,\TRuleConv}
    \frac{
      \derwf \Gamma
      \quad
      \un(\Gamma)
    }{
      \Gamma\der () \colon \Tunit
    }
    \qquad 
    \frac{
      \derwf \Gamma_1,x\colon T,\Gamma_2
      \quad
      \un(\Gamma_1,\Gamma_2) 
    }{
      \Gamma_1,x\colon T,\Gamma_2 \der x \colon T
    }
    \qquad 
    \frac{
      \Gamma_1 \der \varphi\subs vx
      \quad
      \Gamma_2 \der v\colon T
    }{
      \Gamma_1 \csplit \Gamma_2 \der v\colon \{x\colon T \mid \varphi\}
    }
    \qquad
    \frac{
      \Gamma \der v\colon T_1
      \quad
      T_1 \equiv T_2
    }{
      \Gamma \der v\colon T_2
    }
  \end{gather*}
    \emph{Typing rules for processes}, $\Gamma \der P$     
    \begin{gather*}
    \tag{\TRuleEnd,\TRulePar,\TRuleRes,\TRuleRep}
    \frac{
      \derwf \Gamma
      \qquad 
      \un(\Gamma)
    }{
      \Gamma \der \END
    }
    \qquad
    \frac{
      \Gamma_1 \der P_1
      \qquad
      \Gamma_2 \der P_2
    }{
      \Gamma_1 \csplit \Gamma_2 \der P_1 \PAR P_2
    }
    \qquad
    \frac{
      \Gamma \derwf T
      \qquad
      \Gamma,x\colon T,y\colon\dual{T} \der P
    }{
      \Gamma \der \new {xy \tp T} P
    }
    \qquad 
    \frac{
      \un(\Gamma)
      \qquad
      \Gamma \der P
    }{
      \Gamma  \der \REP P    
    }
    \\
    \tag{\TRuleOut}
    \frac{
      \Gamma_1 \vdash x\colon (q\, !y\colon T.U)
      \qquad
      \Gamma_2 \vdash  v \colon T
      \qquad
      \Gamma_3 \cupdate x\colon U\subs vy \vdash P
    }{
      \Gamma_1 \csplit \Gamma_2 \csplit \Gamma_3 \vdash x \out v .P
    }
    \\
    \tag{\TRuleIn}    \frac{
      \Gamma_1 \vdash x\colon (q\,?y\colon T.U)
      \qquad
      (\Gamma_2, z\colon T) \cupdate x\colon U\subs zy \vdash P
    }{
      \Gamma_1 \csplit \Gamma_2 \vdash x \inp z. P
    }
                                    \\
    \tag{\TRuleAssume,\TRuleAssert}
                                \frac{
      \Gamma_2 \der \varphi
      \qquad
      \Gamma_1 \csplit \Gamma_2 \der P
    }{
      \Gamma_1 \der (\assume \varphi ) P    
    }
    \qquad 
    \frac{
      \Gamma_1 \der \varphi
      \qquad
      \Gamma_2 \der P
    }{
      \Gamma_1 \csplit \Gamma_2\der \assert \varphi. P    
    }
    \\
    \tag{\TRuleOneE,\TRuleTensorE,\TRuleRefE}
    \frac{
      \Gamma \der P
    }{
      \Gamma, \one \der P
    }
    \qquad
    \frac{
      \Gamma_1, \varphi_1, \varphi_2,\Gamma_2 \der P
    }{
      \Gamma_1, \varphi_1 \tensor \varphi_2,\Gamma_2 \der P
    }
    \qquad
    \frac{
      \Gamma_1, x\colon T, \varphi\subs xy, \Gamma_2 \der P
    }{
      \Gamma_1, x\colon\{y\colon T\mid \varphi\},\Gamma_2 \der P
    }
  \end{gather*}
  \caption{Typing rules}  \label{fig:type-system}
\end{figure}

We are finally in a position to introduce the type system; the rules
are in Figure~\ref{fig:type-system}. Sequents for extracting formulae
from contexts are of the form $\Gamma \der \varphi$; sequents for
values are of the form $\Gamma \der v\colon T$, and for processes of
the form $\Gamma \der P$.
The rules for formulae should be easy to understand. All our rules
make sure that at the leaves of derivations there are only
well-formed, unrestricted contexts, so as to make sure all linear
entities (formulae and types) are used in a derivation.
The rules for values follow a similar pattern; they include
conventional rules for refinement introduction and for type
conversion.
The first six rules for processes are taken
from~\cite{vasconcelos:fundamental-sessions}.  For instance, the rule
for output splits the incoming context in three parts, one to type the
subject $x$ of communication, the other to type the object $v$, and
the third to type the continuation process $P$. The context for $P$ is
updated with the new type for $x$, that is the continuation type~$U$
with the appropriated substitution applied.

For example, in order to type the final part $s_1 \out 100 . \END$ of
the $\Client_{\textbf 1}$ process under context:
\begin{equation*}
c\colon \ccard, s_1\colon \lin !  a\colon \{ x \colon \nat |
\textit{charge}(c,x) \} . \TEND, \textit{charge}(c,100),
\end{equation*}
 we split the
context in three parts: $c\colon \ccard, s_1\colon \lin !  a\colon \{
x \colon \nat | \textit{charge}(c,x) \} . \TEND$ to type variable
$s_1$, context $c\colon \ccard, \textit{charge}(c,100)$ to type value
100 and context $c\colon \ccard + s_1\colon \TEND\subs{100}{a}$ =
$c\colon \ccard, s_1\colon \TEND$ to type the continuation process
$\END$. From the context for value 100, we build the type $\{ x \colon
\nat | \textit{charge}(c,x) \}$ that matches the ``initial'' part of
the type for $s_1$. Formula $\textit{charge}(c,100)$ is introduced in
the context via the typing rule for assume (see below).

The novelties of the type system are the rules for \keyword{assume}
and \keyword{assert}, and should be easy to understand. Rule
\TRuleAssume adds to the context the formula assumed in the
process. Rule \TRuleAssert works in the opposite direction, removing
from the context the assertion.
Also novel to our type system are the three rule for the elimination
of $\one$, $\otimes$ and refinement types. These rules work in the
context, hence are rules for processes. The corresponding introduction
rules work on the entities (types and formulae) extracted from the
context, and are thus rules for formulae and values.

Back to the running example, let $B_2 = \mu \alpha. \un ! y \colon (
\lin ? c \colon \ccard . \lin ? a \colon \{x \colon \nat |
\textit{charge}(c+10,x) \}.\TEND).\alpha$ be the type of a bank as
seen from the side of the \Store$_{\textbf 1}$ (the type of
$r_2$). Even though we can derive
\begin{multline*}
 s_1 \colon  \lin ! p\colon \product. \lin ! c\colon \ccard . \lin !
  a\colon \{ x \colon \nat | \textit{charge}(c,x) \} . \TEND,
\\
 s_2 \colon  \lin ? p\colon \product. \lin ? c\colon \ccard . \lin ?
  a\colon \{ x \colon \nat | \textit{charge}(c+10,x) \} . \TEND,
  r_2 \colon B_2
  \der \Client \PAR \Store_{\textbf 1}
\end{multline*}
we cannot derive $r_2 \colon B_2 \der \new{s_1s_2}(\Client_{\textbf
  1} \PAR \Store)$ for the types for $s_1$ and $s_2$ are not dual,
because type $\{x \colon \nat | \textit{charge}(c,x) \}$ is not
equivalent to $\{ x \colon \nat | \textit{charge}(c+10,x) \}$, as
required by rule \TRuleRes.

The case of $\Store_{\textbf 2}$ is of a different nature, and in
particular it is not typable due to the impossibility of a suitable
context split. One would like to type $\Store_{\textbf 2}$ under
context:
\begin{equation*}
  s_2\colon \lin ? p\colon \product. \lin ? c\colon
  \ccard . \lin ?  a\colon \{ x \colon \nat | \textit{charge}(c,x) \}
  . \TEND,
  r_2 \colon B_2'
\end{equation*}
where $B_2'$ is type $B_2$ above with $\textit{charge}(c,x)$ replacing
$\textit{charge}(c+10,x)$.
Typing the initial part of the process, using rule \TRuleIn three
times, we introduce in the context the following entries:
$p\colon\product$, $c\colon\ccard$, and $a\colon \{ x \colon \nat |
\textit{charge}(c,x) \}$. Then, using refinement elimination rule,
\TRuleRefE, we convert the last entry in $a\colon\nat,
\textit{charge}(c,a)$. Now, in order to type the continuation
$\Charge(c,x) \PAR \Charge(c,x)$, we have to split the context, but
there is one only formula $\textit{charge}(c,a)$ in the incoming
context, so that only one of the threads will be typable.

On the other hand, consider the case of $\Client_{\textbf 2}$ above
that assumes twice the capability $\textit{charge}(\text c,100)$. By
duality of sessions, the type of the value received by the store will
also be refined with a double capability, $a\colon\{ x \colon \nat |
\textit{charge}(c,x)\tensor \textit{charge}(c,x) \}$. Then we use
\TRuleRefE followed by \TRuleTensorE to obtain $a\colon\nat,
\textit{charge}(c,a), \textit{charge}(c,a)$, making possible the
 split $a\colon\nat, \textit{charge}(c,a) \csplit a\colon\nat,
\textit{charge}(c,a)$.

\section{Main results}
\label{sec:results}

The central result of this paper follows from the lemmas for
weakening, strengthening and
substitution~\cite{vasconcelos:fundamental-sessions} extended to this
system, as well as from basic properties of context splitting (details
omitted).

\begin{lemma}[Weakening]
  \label{lem:weakening}
  If $\Gamma \der P$ and $\un(T)$, then $\Gamma,x:T \der P$.
\end{lemma}

\begin{lemma}[Strengthening]
  \label{lem:strengthening}
  If $\Gamma, x:T \der P$, $\un(T)$ and $x \not\in \fv(P)$, then
  $\Gamma \der P$.
\end{lemma}

\begin{lemma}[Substitution]
  \label{lem:substitution}
  If $\Gamma_1 \der v:T$ and $\Gamma_2,x:T,\Gamma_3 \der P$, then
  $\Gamma_1 \csplit (\Gamma_2,\Gamma_3 \subs vx) \der P \subs vx$.
\end{lemma}

\begin{lemma}[Preservation for $\heat$]
  \label{lem:heating-preservation}
  If $\Gamma\der P$ and $P\heat Q$, then $\Gamma\der Q$.
\end{lemma}

\begin{theorem}[Preservation for $\osred$]
  \label{lem:preservation}
  If $\Gamma\der P$ and $P\osred Q$, then $\Gamma\der Q$.
\end{theorem}

\begin{theorem}[Safety]
  \label{thm:safe}
  If $\Gamma\der P$ and $\un(\Gamma)$, then $P$ is safe.
\end{theorem}

It should be easy to see that processes typable under arbitrary
contexts may not be safe; take for example $A \der \assert A.\END$.

Finally, combining the two results above with a simple induction on
the length of reduction we obtain the main result of the paper.

\begin{corollary}[Main Result]
  \label{thm:main}
  If $\Gamma \der P$ with $\un(\Gamma)$ and $P$ reduces to $Q$ in a
  finite number of steps, then $Q$ is safe.
\end{corollary}

The result states that processes typable under unrestricted contexts
do not get stuck at assertion points (they may still block at input or
output points, due to deadlock).
Furthermore we also know that all assumptions are eventually matched;
e.g., process $(\assume A)\END$ is not typable.
In the case of typable processes it is therefore safe to erase all the
assumptions and assertions from a process, so that there are no
formulae at runtime.

For the cases in proofs involving formulae we make use of the notion
of \emph{canonical contexts}, that is, contexts containing no
refinement types ($x\colon T \in\Gamma$ implies $T$ is not a
refinement type) and whose formulae contain no connectives ($\varphi
\in \Gamma$ implies $\varphi = A$). Contexts can be converted in a
canonical form by using the $\cf$ function, defined on contexts, type
assumptions, and formulae.
\begin{gather*}
  \cf(\cdot) = \cdot
  \qquad
  \cf(\Gamma,\varphi) = \cf(\Gamma),\cf(\varphi)
  \qquad
  \cf(\Gamma,x:T) = \cf(\Gamma), \cf(x:T)
  \\
  \cf(x: \{ y:T | \varphi \}) = \cf(x:T), \cf(\varphi\subs xy)
  \qquad
  \cf(x:T) = x:T \text{ if } T \text{ is not a refinement}
  \\
  \cf(\one) = \cdot
  \qquad
  \cf(\varphi_1 \tensor \varphi_2) = \cf(\varphi_1), \cf(\varphi_2)
  \qquad
  \cf(A) = A
\end{gather*}
We then establish a result, $\Gamma \der P$ iff $\cf(\Gamma) \der P$,
allowing to consider contexts in their canonical form.

\section{Related Work, Conclusions and Future Plans}

Refinements have been useful in verifying polymorphic
contracts~\cite{polycontracts:2011}, security
protocols~\cite{Bhargavan:2010:MVS}, and with the improvements in
satisfiability-modulo-theories (SMT) solvers for classical first-order
logics with uninterpreted functions (such as~\cite{Z3:2008}), can be
integrated into type systems using off-the-shelf components as has
been done for the language F$\#$ using the F7
typechecker~\cite{Fsharp}.

In the context of sessions, Bonelli et al. system of correspondence
assertions for process
synchronization~\cite{bonelli.compagnoni.gunter:correspondence-assertions-process-synchronization}
is close to a basic form of refinement, as it allows labels to be used
in the participant processes of a session to signify the starting and
ending points of marked protocol sections.  This type of protocol
segmentation can be thought of as a basic assume/assert mechanism with
conjunction, since multisets of labels can be used for the part
equivalent to `assert,' but still without the rich constructors and
proof system of a logic.
  Bocchi et al.\ introduce assertions in multiparty session types
(session types allowing to describe interaction among multiple
partners)~\cite{bocchi.honda:theory-of-design-by-contract}. Similarly
to the system of Bonelli et
al.~\cite{bonelli.compagnoni.gunter:correspondence-assertions-process-synchronization},
assertions are explicitly associated with session operations (in, out,
branch, select). In contrast, our system introduces assertions as
refinement types to be used at arbitrary places in a protocol;
furthermore their system uses classical logic as opposed to linear
logic.

The recent work by Toninho et
al.~\cite{toninho.caires.pfenning:dependent-session-types} interprets
session types within intuitionistic linear logic, obtaining (with some
extensions) a dependent sessions type system for $\pi$-calculus.
This system interprets session types as linear logic formulae, with
input as $\multimap$ and output as $\otimes$, and stratifies the
language into a $\pi$-calculus for communication and a functional
language for proof objects, where the latter are opaque terms that (in
our system) would correspond to proofs of refinements.  However, their
system does not consider \emph{linear} refinements, i.e., linearity is
restricted to the communication layer (the sessions).
Although the aims of both systems are similar to an extent, we have
taken a different approach, adopting session types without their
linear-formulae interpretation, and focussing on the incorporation of
fine-grained linear refinements which provide for a more delicate
distinction between types. Moreover, we do not utilise proof-witnesses
but rather implement proof search within the type system itself; then,
using the heating relation, assumptions are manipulated
at runtime in order to check assertions, which is essentially a
procedure of cut-elimination.

The concept of names appearing in types was pioneered in the work by
Yoshida on channel-dependent types for processes with code
mobility~\cite{Yoshida:2004:CDT}, and was adapted to sessions in
subsequent
work~\cite{mostrous.yoshida:two-session-systems-higher-order}.  In
these systems there are no refinements, yet channel dependent types
are shown to provide security guarantees by controlling which names
may be used in communications between received code and host
environment, which indicates that an integration with our system could
provide even greater control over mobile code.

In summary, a theory of (linear) refinement types for sessions has not
been hitherto proposed, marking the contribution of our system.
As future work, it is interesting to consider sessions \emph{as}
linear refinements, and to extend our refinement language to a larger
fragment of Linear Logic.  We plan to investigate decidable
type-checking, drawing inspiration from the techniques
in~\cite{Flanagan:2006:HTC,Rondon:2008:LT,vasconcelos:fundamental-sessions},
and by considering appropriate restrictions.  Moreover, it would be
interesting to examine the adaptations necessary for languages with
(asynchronous) buffered semantics, where communications can be
reordered, especially in the context of mobile session-typed
processes~\cite{mostrous_yoshida_tlca09}, channel dependent
types~\cite{mostrous.yoshida:two-session-systems-higher-order,
  Yoshida:2004:CDT}, and multi-party sessions~\cite{Honda:2008:MAS}.

\textbf{Acknowledgements.}  This work was supported by projects
Interfaces, CMU-PT/NGN/0044/2008, and Assertion-types,
PTDC/EIA-CCO/105359/2008.

\end{document}